\documentclass[12pt]{article}
\usepackage{latexsym, longtable}
\usepackage{amssymb, amsmath, xspace, lscape,  latexsym}

\renewcommand{\le}{\leqslant}

\def\beq#1#2\eeq{%
        \begin{equation}%
        \label{#1}%
            #2%
        \end{equation}%
    }

\newcommand{\lb}[1]{\label{#1}}

\newcommand{\mref}[1]{(\ref{#1})}

\newcommand{\p}{\partial}

\renewcommand{\a}{\alpha}
\renewcommand{\b}{\beta}

\newcommand{\R}{\mathbb R}

\renewcommand{\hat}{\widehat}
\renewcommand{\tilde}{\widetilde}

\def\btheor#1\etheor{%
        \begin{theor}%
            #1%
        \end{theor}%
    }

    \def\bsled#1\esled{%
        \begin{sled}%
            #1%
        \end{sled}%
    }

\newtheorem{theorem}{Theorem}

\newtheorem{lemma}{Lemma}
\newtheorem{prop}{Proposition}
\newtheorem{cor}{Corollary}

\newcommand{\nad}[2]{\genfrac{}{}{0pt}{}{#1}{#2}}

\def\hm#1{#1\nobreak\discretionary{}{\hbox{\m@th$#1$}}{}}
\def\mi#1{\discretionary{\hbox{\m@th$#1$}}{\hbox{\m@th$#1$}}{}}

\vspace{4ex}
\begin{document}

\begin{center}

{\bf\Large Logarithmic Frobenius structures and Coxeter discriminants}

\vspace{1cm}

{\bf  M. V. Feigin$^1$ and A. P. Veselov$^{2}$}
\end{center}

\vspace{1cm} \noindent $^1$ Department of Mathematics, University of
Glasgow, University Gardens, Glasgow G12 8QW, UK

\noindent $^2$
Department of Mathematical Sciences,
Loughborough University,\\
Loughborough,  LE11 3TU, UK
and  Landau Institute for Theoretical Physics,
Kosygina 2, Moscow, 117940, Russia

\noindent E-mail addresses: m.feigin@maths.gla.ac.uk,
A.P.Veselov@lboro.ac.uk

\vspace{1cm}

\begin{abstract}
\noindent We consider a class of solutions of the WDVV equation
related to the special systems of covectors (called $\vee$-systems)
and show that the corresponding logarithmic Frobenius structures can
be naturally restricted to any intersection of the corresponding
hyperplanes. For the Coxeter arrangements the corresponding
structures are shown to be almost dual in Dubrovin's sense to the
Frobenius structures on the strata in the discriminants discussed by
Strachan. For the classical Coxeter root systems this leads to the
families of $\vee$-systems from the earlier work by Chalykh and
Veselov. For the exceptional Coxeter root systems we give the
complete list of the corresponding $\vee$-systems. We present also
some new families of  $\vee$-systems, which can not be obtained in
such a way from the Coxeter root systems.
\end{abstract}

\vspace{1cm}


\section{Introduction}

The space of orbits  $M_G$ of a finite Coxeter group $G$ is probably
the most remarkable example of the Frobenius manifolds \cite{D, D1,Saito}.
In fact it is a unique in some sense according to the Dubrovin
conjecture \cite{D} proved by Hertling \cite {H}. It has the only
disadvantage that in general the corresponding prepotential (known to be polynomial)
cannot be written explicitly in a simple way.

On the other hand one can show that for any Coxeter root system $R \subset V$ the function
\beq{FR}
F_R(x) = \sum_{\a\in R}(\a,x)^2 \log (\a,x)^2
\eeq
satisfies the so-called WDVV equations (see the next section) and thus determines some multiplication structure on the tangent bundle of the complement $\Sigma_R$ to the mirror hyperplanes $(\alpha,x)=0$ (see
\cite{MG, V1}). This type of solutions for the WDVV equation came from the Seiberg-Witten investigations of  $N=2$ SUSY Yang-Mills theory \cite{MMM}.

It turned out that in this respect the Coxeter root systems are not
unique: in \cite{V1,V2} it was shown that the deformed root systems
discovered by Chalykh and the authors in \cite{VFC, CFV} also give
the solutions of the WDVV equations. This led to the notion of the
$\vee$-systems \cite{V1}, which can be considered as a proper
extension of the Coxeter root systems for this problem. Some new
families of the $\vee$-systems generalising classical Coxeter cases
were found later in \cite{CV}.

The answer to a natural question what is the relation between the two types of Frobenius structures related to Coxeter groups was found recently by Dubrovin \cite{D2}. He showed that the original Frobenius structure on $M_G$ and the corresponding product on $\Sigma_R$ (denoted by star) are related by the following remarkably simple formula:
\beq{twoprodin}
u * v = E^{-1} \cdot u \cdot v,
\eeq
where $E$ is the Euler vector field.

In this paper we show that this relation  \mref{twoprodin} between two multiplications can be extended to any stratum of the {\it Coxeter discriminant} $\Sigma \subset M_G.$ The restrictions of the Frobenius structure on $M_G$ to the strata of the discriminant were considered first by Strachan \cite{S}, who investigated the natural submanifolds of the Frobenius manifolds.

Thus we extend Dubrovin's duality to the duality between Strachan's structures on the natural submanifolds and logarithmic Frobenius structures with the prepotentials of the form \mref{FR}, where $R$ is to be replaced by certain $\vee$-systems.
These systems can be described as projections (or restrictions in the dual picture) of the Coxeter root systems to the corresponding intersection subspace of the mirror hyperplanes.

Remarkably enough in this way we get a geometric explanation of the
$\vee$-systems found in \cite{CV}, which can be considered therefore
as a closure (in Zariski sense) of the discrete infinite set of the
restricted classical Coxeter root systems.

We use the results of Orlik and Solomon \cite{OS} and Shcherbak
\cite{Shch}, who classified the strata in the Coxeter discriminant,
to give the complete list of the $\vee$-systems, which are the
restrictions of the exceptional Coxeter root systems. The case of
the root system of type $F_4$ leads to new interesting examples of
the $\vee$-systems, which we denote $F_3, F_5, F_6$ (see section 5
below).

Finally, we present some new families of the $\vee$-systems, which
can not be obtained as the restrictions of the Coxeter root systems.

\section{WDVV equation, $\vee$-systems and logarithmic Frobenius structures}

The (generalised) WDVV equation is
the following overdetermined system of nonlinear partial differential equations in $\bf{R}^n$:
\beq{wdvv}
F_iF_k^{-1}F_j=F_jF_k^{-1}F_i, \quad i,j,k=1,\ldots,n,
\eeq
where $F_m$ is the $n\times n$ matrix constructed from the third partial
 derivatives of the unknown function $F=F(x^1,\ldots,x^n)$:
\beq{f}
(F_m)_{pq}=\frac{\partial ^3 \, F}{\partial x^m \partial x^p \partial x^q}.
\eeq
Let us introduce for any vector field $a = a^i \partial_i$ the
matrices $ F_{a}= a^i  F_i$ (here and below the summation over repeated indices is assumed).
It is known \cite{MMM}  that (\ref{wdvv}) are equivalent to the equations
\beq{fgf}
F_iG^{-1}F_j=F_jG^{-1}F_i, \quad i,j=1,\ldots ,n,
\eeq
where $G=F_{\eta}$ for some vector field $\eta$, which is assumed to be invertible.
It is easy
to see that one can rewrite (\ref{fgf}) as the commutativity
relations \beq{com} \left[ \hat F_a, \hat F_b \right] =0 \eeq where $\hat F_a = G^{-1} F_a, \hat F_b = G^{-1} F_b$ for any two vector fields $a$ and $b$.

Consider the
(pseudo-Riemannian) metric  $$<a,b> =  G_{ij} a^i  b^j = F_{ijk}
a^i  b^j \eta^k$$  and define  the multiplication on the tangent
bundle by the formula \beq{multip} a*b = \hat F_a (b) = \hat F_b
(a). \eeq It has the following properties:
\begin{enumerate}
\item commutativity: $a*b = b*a$
\item associativity: $(a*b)*c = a*(b*c)$
\item Frobenius property: $<a*b, c> = <a, b*c>.$

\end{enumerate}

The first and the last properties immediately follow from the symmetry of the partial derivatives,
but the associativity imposes a non-trivial condition on $F$, which is nothing else but the WDVV relation (\ref{wdvv}),(\ref{com}). Comparing all this with Dubrovin's definition of the Frobenius manifold \cite{D2} we see that we lack some properties: in general metric $G$ is not flat and the vector field $\eta$ is not covariantly constant.

Let us consider now the following particular class of the solutions
of the WDVV equation \cite{V2}. Let $V$ be a real linear vector
space of dimension $n$, $V^*$ be its dual space consisting of the
linear functions on $V$ (covectors),  ${A}$ be a finite
 set of covectors $\alpha \in V^*$. One can always assume them to be noncollinear, although sometime it is convenient not to do this (see below).

Consider the following function on $V$: \beq{mF}
F^{{A}}=\sum\limits_{\alpha \in {A}} (\alpha,x)^2 \, {\rm log} \,
(\alpha,x)^2, \eeq where $(\alpha,x)=\alpha(x)$ is the value of
covector $\alpha \in V^*$ on a vector $x\in V$. It is defined on the
complement $\Sigma_{{A}} = V \setminus \cup_{\alpha \in {{A}}}
\Pi_{\alpha}$ to the union of all hyperplanes $\Pi: \alpha(x)=0$.
One can check that the corresponding $F_a$  is (up to a constant)
the matrix of the following bilinear form on  $V$ $$
F_a^{{A}}=\sum\limits_{\alpha \in {A}} \frac{(\alpha,a)}{(\alpha,x)}
\alpha\otimes \alpha, $$ where $\alpha\otimes \beta (u,v)=\alpha
(u)\beta (v)$ for any $u,v \in V$ and $\alpha, \beta \in V^*$.

If we choose the vector field $\eta$ to be the Euler vector field $\eta = x^i \partial_i$ we come to the constant matrix $G^{{A}} = F_x^{{A}},$
corresponding to the following bilinear form
\beq{mG}
G^{{A}}=\sum\limits_{\alpha \in {A}} \alpha\otimes \alpha.
\eeq

We will assume now that this form is non-degenerate, which in the
real case means that the covectors $\alpha \in {A}$ generate
$V^*$. Then the natural linear map $\varphi_{A} : V\rightarrow
V^*$  defined by the formula $$ (\varphi_{A}(u),v)=G^{{A}} (u,v),
\, u,v \in V $$ is invertible. We will denote
$\varphi_{A}^{-1}(\alpha),\,  \alpha \in V^*$ as $\alpha^{\vee}$.
By definition the operator $$\sum\limits_{\alpha \in {A}}
\alpha^{\vee}\otimes \alpha = Id$$ is an identity operator in $V$,
or equivalently \beq{vee} (\alpha,v)=\sum\limits_{\beta \in
{A}}(\alpha, \beta^{\vee})(\beta,v) \eeq for any $\alpha \in V^*,
v \in V$. In this notation the operators $\hat F_a^{{A}}$ can be
written as $$ \hat F_a^{{A}}=\sum\limits_{\alpha \in {A}}
\frac{(\alpha,a)} {(\alpha,x)}\alpha^\vee \otimes \alpha. $$

This leads to the following multiplication for the tangent vectors $u$ and $v$
on $\Sigma_{{A}}$:
\beq{mult}
u * v = \sum_{\a\in {{A}}} \frac{\a(u) \a(v)}{\a(x)} {\a^\vee}
\eeq

A simple calculation  \cite{V2} shows that the associativity of this multiplication (which is the same as WDVV relation for (\ref{mF}) or commutativity of $\hat F_a^{{A}}$) can be rewritten as
\beq{sc}
\sum\limits_{\alpha \ne \beta, \alpha,\beta \in {A}}
\frac{G^{A} (\alpha^\vee, \beta^\vee)B_{\alpha,\beta}(a,b)}
{(\alpha,x)(\beta,x)}\alpha\wedge \beta \equiv 0,
\eeq
where
$$
\alpha\wedge \beta=\alpha\otimes \beta-\beta \otimes \alpha
$$
and
$$
B_{\alpha,\beta}(a,b)=\alpha\wedge \beta(a,b)=\alpha(a)\beta(b)-\alpha(b)\beta(a).
$$
Another interpretation of these relations is
the commutativity condition of the
following differential operators of the Knizhnik-Zamolodchikov type
\beq{dkz}
\bigtriangledown_a = \partial_a - \sum\limits_{\alpha \in {A}}
\frac{(\alpha,a)}{(\alpha,x)}\alpha^\vee \otimes \alpha,
\eeq
which therefore define a flat connection
on $\Sigma_{{A}}$ (see \cite{V2}).

The corresponding sets ${{A}},$ for which all these equivalent properties hold, are called $\vee$-{\it systems} \cite{V1}. They satisfy the following relations, called $\vee$-{\it conditions}:
\beq{expl}
\sum\limits_{\beta \in \Pi \cap {A}}
\beta(\alpha^\vee)\beta^\vee=\lambda \alpha^{\vee},
\eeq
for any two-dimensional plane $\Pi \subset V^*, \alpha \in \Pi \cap {A}$ and some $\lambda$, which may depend on $\Pi$ and $\alpha$ (see \cite{V2}).

To give a more geometric description let us introduce the
Euclidean structure on $V$ using the form $G^{{A}}$. We say that a
finite set ${{A}}$ in the Euclidean vector space is {\it
well-distributed} if $$\sum\limits_{\alpha \in {A}}(\alpha, x)
(\alpha, y) = \lambda (x,y)$$ for some $\lambda$ and {\it
reducible} if ${A} = {A}_1 \cup {A}_2$ is a union of two non-empty
orthogonal subsystems. Then ${A}$ is a $\vee$-system if it is
well-distributed and any its two-dimensional subsystem is either
reducible or well-distributed in the corresponding plane (see
\cite{V2}).

For any $\vee$-system ${A}$ the formula (\ref{mult})
defines what we call {\it logarithmic Frobenius
structure on $\Sigma_{{A}}$ with prepotential} (\ref{mF}). It
satisfies all the properties of the Frobenius structure in
Dubrovin's sense \cite{D2} except the covariant constancy of the
unit vector field. In particular, logarithmic Frobenius structure
defines an $F$ manifold with flat structure compatible with the
multiplication in the sense of \cite{M}.

 The full classification of the
$\vee$-systems is still to be done. Here are the examples known so
far:

1) Any two-dimensional system (trivial examples)

2) Coxeter $\vee$-systems \cite{MG, V1}

3) Deformed $A_n$ and $B_n$ families \cite{CV}

4) Deformed root systems related to simple Lie superalgebras
\cite{SV1}.



By a {\it Coxeter root system} $\mathcal R$ we will mean a finite
set of non-zero vectors in Euclidean space such that for any root
$\alpha \in \mathcal R$ the reflection $s_{\alpha}: x \rightarrow  x
-\frac{2(\alpha,x)}{(\alpha, \alpha)} \alpha$ leaves $\mathcal R$
invariant: $s_{\alpha}\mathcal R = \mathcal R.$ We assume also that
the only root from $\mathcal R$ collinear to $\alpha \in \mathcal R$
apart from $\alpha$ itself is $-\alpha.$ The reflections
$s_{\alpha}, \alpha \in \mathcal R$ generate a finite {\it Coxeter
group} $G.$

In other words the Coxeter root systems consist of the normals to
the mirror hyperplanes (two for each mirror), normalised in such a
way that the length is a constant on each orbit of the corresponding
Coxeter group. Note that these systems depend on the parameters
whose number is equal to the number of orbits of the group on its
root system (or one less if we consider these systems up to a
dilation).

For a given Coxeter root system $\mathcal R$ we define the
corresponding $\vee$-system $A \subset \mathcal R$ as a subset of
non-collinear roots such that $\mathcal R = A \cup (-A).$ The
standard choice is so-called {\it positive part} of root system
$\mathcal R_{+},$ consisting of vectors positive with respect to
some linear form.

The deformed $A_n$ system has the form $$
{A}_n(c)=\{\sqrt{c_ic_j}(e_i-e_j),\quad 1\le i<j\le n+1\}$$ in $\mathbb
R^{n+1},$ where $c_1,\dots, c_{n+1}$ are arbitrary
parameters. The deformed $B_n$ family depends on $n+1$ parameters $(\gamma; c_1, \dots, c_n)$  and has the form
\begin{equation}\label{bc}
{B}_n(\gamma, c)= \left\{
\begin{array}{lll}
\sqrt{c_ic_j}(e_i \pm e_j)\,, &  1\le i<j\le n\,,
\\ \sqrt{2 c_i(c_i + \gamma)}e_i\,, &  i=1,\ldots ,n\,
\end{array}
\right.
\end{equation}
(see \cite{CV}). The construction of the next section gives some
explanation of these two families (see also section \ref{coxres}).

\section{ Restrictions of the logarithmic Frobenius structures on the intersection subspaces}

Let ${A} = \{\a\} \subset V^*$ be a $\vee$-system, $F(x) = \sum_{\a
\in {A}} \a(x)^2 \log \a(x)^2$ be the prepotential of the corresponding
logarithmic Frobenius structure.

By a {\it subsystem} ${B} \subset {A}$ we will mean the intersection
of ${A}$ with any subspace $U \subset V^*.$ Choose any subsystem
${B}$ in $ {A}$ and consider the corresponding subspace $L = L_B
\subset V,$ which is the intersection of the hyperplanes $\Pi_{\b} =
\{ x \in V: \b(x)=0\}$ for all $\b \in {B}$. On the first sight the
prepotential $F$ and the corresponding multiplication (\ref{mult}) can
not be restricted to $L,$ but in fact this can be done in the following natural way.

Let $C = A \setminus B$ be the complement to $B$ in $A.$
Consider the corresponding space $L \setminus \Sigma_L$, which is a complement in $L$ to the union $\Sigma_L$ of the intersection hyperplanes $\tilde{\Pi}_{\gamma} = \Pi_{\gamma} \cap L$ for all $\gamma \in C.$

\begin{theorem}\label{thm1}
The logarithmic Frobenius structure (\ref{mF}), (\ref{mult}) has a natural restriction to the space $L  \setminus \Sigma_L$
with the prepotential
\beq{potr}
F_B = \sum_{\gamma\in A \backslash B} \gamma(x)^2 \log \gamma(x)^2, \quad x \in L  \setminus \Sigma_L,
\eeq
which also satisfies the WDVV equation.
\end{theorem}

\begin{cor}\label{cor1}
The restriction of a $\vee$-system $A$ to the intersection subspace $L = L_B$
for any subsystem $B \subset A$ is a $\vee$-system as well.
\end{cor}

Note that if we would assume the covectors in the original system $A$ to be noncollinear,
the same will not be true in general for the restricted system. This is however not a problem since
a group of the collinear covectors $\gamma_i = \lambda_i \gamma, i=1,\dots, k$ can be replaced by a single covector $\tilde \gamma = \lambda \gamma,$ where $\lambda^2 = \sum_{i=1}^k \lambda_i^2.$


Let us now prove the theorem. Consider a point $x_0\in \Sigma_L$ and two tangent vectors $u,v$ at $x_0$ to $\Sigma_L.$
Let us extend vectors $u$ and $v$ to two local analytic vector
fields $u(x), v(x)$ in the whole space $V$  tangent to the subspace
$L.$  Outside $L$ we have a well-defined multiplication $u(x)*v(x),$
so the question is what happens when we approach $x_0 \in L.$ The
answer is given by the following

\begin{lemma}\label{lem0}
The product $u(x) * v(x)$ has a limit when $x$ tends to $x_0$
given by
$$
u * v = \sum_{\a \in A\backslash B} \frac{\a(u) \a(v)}{\a(x_0)}
{\a^\vee}.
$$
In particular, the limit is determined by $u$ and $v$ only.
\end{lemma}
{\bf Proof.} It is enough to analyse the singular part of the
product $u(x) * v(x)$ near a hyperplane $\Pi_\b: \b(x)=0, \, \b \in
B$. Consider Euclidean local coordinates $(t, s)$ near the
hyperplane such that $t=\beta(x)$ is the coordinate along the normal
direction, and $s$ is vector of $n-1$ coordinates of the orthogonal
projection of $x$ onto $\Pi_\beta$. The vector fields can be written
as
$$
u(x)=u(t, s)= a(t, s) \partial_t + \xi(t, s),
$$
$$
v(x)=v(t, s)= b(t, s) \partial_t + \eta(t, s),
$$
where vector fields $\xi$ and $\eta$ are parallel to the hyperplane
$\Pi_\beta$: $\beta (\xi) = \beta (\eta) = 0.$ Since the fields are
assumed to be tangential to $\Pi_\beta$ we have $a(0, s)=b(0, s)=0$.
 The coefficients $a,b$ are analytic, so
$$
\lim_{t \to 0} \frac{a(t, s) b(t,  s)}{t} =0.
$$
This means that $u(x)*v(x)$ is non-singular at $\b(x)=0$, and that the $\b$
term disappears when calculating the product at $\Pi_\b$. As $\b$ is an arbitrary
element from the subset $B$ the lemma follows.

Thus the $*$-product is defined for two tangent vectors to $\Sigma_L.$ The next question is whether it belongs to the tangent space to  $\Sigma_L.$  This is true and follows immediately from the following statement.

\begin{lemma}\label{lem1}
Let $A$ be a $\vee$-system, $\a \in A$ be its element,
$\Pi_\a=\{x|\a(x)=0\}$ be the corresponding hyperplane. Then if
$u,v$ are tangent to $\Pi_\a$  then the same is true for $u * v.$
\end{lemma}
{\bf Proof.}
We have to show that
$$
\sum_{\nad{\b\in A}{\b\ne\a}} \frac{\b(u) \b(v)}{\b(x)} \a(\b^\vee) = 0
$$
if $x\in \Pi_\a; u,v \in T_x\Pi_\a$. From the relation (\ref{sc}),
which is true for any $\vee$-system, we have the following identity
$$
\sum_{\nad{\b\in A}{\b\ne\a}} \frac{G^{A} (\alpha^\vee, \beta^\vee)(\a(a)\b(b)-\a(b)\b(a))(\a(z)\b(y)-\a(y)\b(z))}{\b(x)}
\equiv 0
$$
holding on the hyperplane $\a(x)=0$, where
$a,b,y,z$ are arbitrary vectors in $V$.
We take $b=u \in \Pi_\a$,  $y=v \in  \Pi_\a$. Then $\a(b)=\a(y)=0$ and
we get
$$
\a(a) \a(z) \sum_{\nad{\b\in A}{\b\ne\a}} \frac{\a(\b^\vee)\b(u) \b(v)}{\b(x)} = 0,
$$
which implies the lemma.

The fact that the restricted structure has the prepotential (\ref{potr}) is a simple check now.
The theorem is proved.

We should mention that a related result was recently found also by
Couwenberg, Heckman and Looijenga \cite{CHL}. In fact, the main
object of their paper (Dunkl system) is very close to the notion of
the $\vee$-system.

In the next two sections we  discuss the restricted logarithmic
Frobenius structures for the Coxeter root systems.

\section{Dubrovin's duality on the strata of the Coxeter discriminant}

Let $G$ be a finite Coxeter group generated by reflections in
$V=\R^n$ and $M_G = V / G$ be the corresponding space of orbits. Let
$y_1, \ldots, y_n$ be a set of free homogeneous generators of the
algebra of the $G$-invariant polynomials with degrees $d_1, d_2,
d_3,\ldots, d_n=2$ assumed to be in the decreasing order. It is
known that $M_G$ can be equipped with the Frobenius manifold
structure \cite{D1}. The identity vector field is $e=\frac{\p}{\p
y_1}$, the Euler vector field is proportional to
$$
E=\sum_{i=1}^n d_i y_i \frac{\p}{\p y_i}.
$$

Dubrovin introduced the following notion of almost dual Frobenius
manifold \cite{D2}. For a given Frobenius manifold the dual $\star$
product is defined as follows
\beq{twoprod}
u \star v = E^{-1} \cdot u \cdot v
\eeq
on the set where the operator of multiplication by $E$ is invertible.

Dubrovin has also shown that for the Coxeter orbit space $M_G$
the dual structure is in fact the
logarithmic Frobenius structure with the prepotential \beq{F} F(x) =
 \sum_{\a\in R} (\a,x)^2 \log (\a,x)^2, \eeq where
$R$ is the set of normals (of the same length) to the mirrors $(\a,x)=0$ of the group
$G.$

Now we are going to show that Dubrovin duality relation
\mref{twoprod} can be extended to any strata of the Coxeter
discriminant. The restrictions of the Frobenius structure on $M_G$
to the strata of the discriminant variety were considered by
Strachan \cite{S}. We will show now that Dubrovin's duality can be
extended to Strachan's structures and the corresponding dual
structures are nothing else but the logarithmic Frobenius structures
related to the $\vee$-systems, which are the restrictions of the
Coxeter root systems.

To make this precise recall that the Coxeter discriminant $\Sigma \subset M_G$
consists of irregular orbits, which are orbits of $G$ consisting of
less than $|G|$ points. It is the image of the union of the
reflection hyperplanes under the natural projection map $\pi: V \to M_G.$ Consider a
stratum $S = \pi (L)$ in $\Sigma,$ which is the image of the
intersection subspace $L = \cap_{\b\in B} \Pi_\b,$ $B$ is a
subsystem in the root system $R$ of $G$.

According to Strachan \cite {S} $S$ is a {\it natural submanifold} of the Frobenius
manifold $M_G,$ which means that the tangent space at any regular point of $S$
is closed under multiplication and contains the Euler vector field.
The restricted structure does not satisfy Dubrovin's axioms \cite{D1} of the Frobenius manifold,
since the corresponding metric is not flat anymore, see \cite{S}.

Consider a point $x \in L \setminus  \Sigma_L,$  where as before
$\Sigma_L$ is the union of the hyperplanes $\tilde{\Pi}_{\gamma} =
\Pi_{\gamma} \cap L, \quad \gamma \in R \setminus B.$ Let $u,v$ be
two vectors tangent to $L$ at $x$. Since the natural projection
$\pi$ restricted to $L$ is a local diffeomorphism at $x$ we can
consider $u, v$ also as tangent vectors at $y = \pi(x) \in S.$ We
know that the logarithmic Frobenius structure with prepotential
(\ref{F}) can be restricted to $L$ and the corresponding
prepotential on $L  \setminus \Sigma_L$ is given by \beq{potres} F_B
= \sum_{\gamma\in R \backslash B} \gamma(x)^2 \log \gamma(x)^2, \eeq
where $\gamma(x)=(\gamma,x)$. The following theorem is a corollary
of the previous results and the results of Dubrovin \cite{D2} and
Strachan \cite{S}.

\begin{theorem}\label{dualitystrat}
The restriction of the Frobenius structure on $M_G$ to the stratum $S$ in the Coxeter discriminant $\Sigma$ and the logarithmic Frobenius structure on $L \setminus \Sigma_L$ with
prepotential (\ref{potres}) are related by Dubrovin's duality formula
\beq{DD}
u * v = E^{-1} \cdot u \cdot v.
\eeq
\end{theorem}

We would like to note that all the roots in $R$ in this claim are
normalised to have the same length according to Dubrovin's duality
theorem. It is not clear what kind of structure on $M_G$ corresponds
to the case when the lengths depend on the choice of orbit.

\section{Restrictions of the Coxeter root systems}\label{coxres}

In this section we discuss $\vee$-systems which can be obtained as
the restrictions of the Coxeter $\vee$-systems.

\subsection{$A_n$-type systems}

We start with the standard $A_n$ root system consisting of the covectors $e_i - e_j$ with $i, j = 1, 2, \dots, n+1$.

Let $c = (c_1, c_2, \dots, c_m) $ be any partition of $n+1$:
$$\sum_{i=1}^m c_i = n+1, c_1 \geq c_2 \geq \dots \geq c_m \geq 1.$$
It defines a natural representation of the set $I_{n+1} = \{ 1, 2, \dots, n+1\}$ as a union of the subsets
$I_{n+1} = \bigcup_{k=1}^{m} C_k$ with $|C_k| = c_k.$
The set $B_c \subset A_n$ of the covectors $e_i - e_j$, where $i$ and $j$ belong to the same subset $C_k$ is a subsystem of $A_n$ and any subsystem can be represented in such a way modulo permutation, which is the action of the corresponding group $G=S_{n+1}.$

The corresponding subspace $L_c$ is given by the condition that all the coordinates with indices inside the same group $C_k, \, k=1,\dots, m$ are equal. The restrictions of the covectors $e_1, e_2, \dots, e_{c_1}$ to $L_c$ are the same; let us denote the corresponding covector $f_1.$ Similarly we define $f_k, \, k=2, \dots, m$ for all other groups.
In these notation the restriced system
consists of the covectors $f_i-f_j$ with multiplicities $c_i c_j$. Therefore we get the following

\begin{prop}\lb{propan}
The $\vee$-systems, which can be obtained by the restriction of the $A_n$ system,
have the form
\beq{anc}
\a_{ij}=\sqrt{c_i c_j} (f_i-f_j),
\eeq
where $1\le i < j \le m\le n+1$ and  $c = (c_1, c_2, \dots, c_m) $ is a partition of $n+1$.
\end{prop}

Note that the system \mref{anc} is a $\vee$-system for general, not necessarily integer values of the parameters $c_i.$
The corresponding solution to the WDVV equation was first found in the paper \cite{CV}, where it was also proved that for $n=3$ this is the most general solution of $A$-type.

{\bf Remark.} The extension of the values of the parameters from
integers to the real (complex) is actually automatic. Indeed the
$\vee$-conditions are equivalent to the set of algebraic relations
on the parameters of the solution. If they are satisfied for all
integers they must be valid for all values of the parameters. In
particular, the same is true for the families coming from all the
classical Coxeter root systems.

\subsection{$BCD_n$-type systems}

The Coxeter group of $B_n$-type has two orbits on its root system,
which leads to an extra parameter in our construction, which we will
denote $\Lambda.$ Consider the following set of covectors in $\R^n$:
\beq{lambdacl} e_i \pm e_j, \,  \Lambda e_i, \eeq for $1\le i<j \le
n$. When $\Lambda=0,1,2$ one gets the positive part of the root systems $D_n, B_n, C_n$
respectively. The set defines a $\vee$-system for any value of the parameter
$\Lambda$ ( see \cite{MG, V1}), which we will denote as $B_n(\Lambda).$

Let $c_0$ be an integer from the set $\{0, 1, \dots, n\}$ and $c = \{c_1, c_2, \dots, c_m\}$ be a partition of $n-c_0$: $c_1 + \dots + c_m = n - c_0, \quad  c_1 \geq c_2 \geq \dots \geq c_m \geq 1.$

Let $I_{n} = \bigcup_{k=0}^{m} C_k, \quad |C_k| = c_k$ be the corresponding partition of the set $I_{n} = \{ 0, 1, \dots, n\}.$ Consider the subsystem consisting of the covectors $\Lambda e_s, \, s \in C_0$ and $e_i - e_j,$ where $i$ and $j$ belong to the same subset $C_k, \, k=1,\dots, m.$ This is the most general subsystem in the $B_n, C_n$ case modulo action of the corresponding group $G.$ In the $D_n$-case there are other types but they lead to the same $\vee$-systems. We should add also that $c_0 \neq 1$ in the $D_n$-case.

The corresponding subspace $L_c$ is defined by the condition that the first $c_0$ coordinates are equal to zero and the coordinates with the indices within the same subset $C_k, \, k=1,\dots, m$ are equal.
All the covectors from a subset $C_k$ have the same restriction to $L_c$, which we will denote as $f_k, k=1, \dots, m.$ The restriction gives the following set of covectors with
multiplicities:
\begin{eqnarray*}
f_i \pm f_j, & \mbox{multiplicity} & c_i c_j,
\\
f_i, & \mbox{multiplicity} & 2 c_i c_0,
\\
2 f_i,&   \mbox{multiplicity} & \frac{c_i(c_i-1)}2,
\\
\Lambda f_i, & \mbox{multiplicity}& c_i,
\end{eqnarray*}
where $1 \le i<j \le m$. Equivalently, in the resulting $\vee$-system we
may change all the covectors proportional to $f_i$ to just one covector
$$
\sqrt{2 c_i^2 + c_i (\Lambda^2 + 2 c_0 -2)} f_i.
$$

\begin{prop}
The $\vee$-systems, which are the restrictions of the $B_n(\Lambda)$
system \mref{lambdacl}, have the form \beq{anc2} \sqrt{c_i c_j} (f_i
\pm f_j), \sqrt{2 c_i (c_i + \gamma)} f_i \quad (1\le i < j \le m)\eeq
where $ \gamma = \frac{1}{2} (\Lambda^2 + 2 c_0 - 2),  \quad  \sum_{i=0}^m c_i =n$ with integer $c_0 \geq 0, \quad  c_1 \geq c_2 \geq \dots \geq c_m \geq 1.$
\end{prop}
These $\vee$-systems (for general values of the parameters) were found in \cite{CV}. We denote them $B_m(\gamma; c_1, \dots , c_m).$ As before the Coxeter restrictions corresponding to the integer values of the parameters form a subset, which is dense in Zariski sense.

\subsection{Exceptional systems: $ \pmb \vee$-systems of $F_n$-type}
\label{exceptF}

In this section we discuss the restrictions of the root system $F_4$ and the analogues of the corresponding $\vee$-systems in higher dimensions.

The Coxeter $\vee$-system of $F_4$-type is the following set in
$\R^4$: \beq{F4}
 e_i \pm e_j, 2 \Lambda e_i, \, \Lambda(e_1\pm e_2 \pm e_3 \pm e_4),
\eeq
where $1 \le i < j \le 4$,  $\Lambda \in \R$, and the signs can be chosen
arbitrarily.
Again an additional parameter $\Lambda$ is due to the existence of two orbits of the corresponding group $G = F_4$ on its root system.

There are two non-equivalent choices of the one-dimensional
subsystems: $B = \{2 \Lambda
e_1\}$ and $B = \{e_1-e_2\},$ leading to the following $\vee$-systems. In the first case we get the system $F_3^1(\Lambda)$:
\begin{multline}
\label{f41}
e_1 \pm e_2, e_2 \pm e_3, e_1 \pm e_3, \sqrt{4\Lambda^2+2}\,e_1, \sqrt{4\Lambda^2+2}\,e_2,
\\
\sqrt{4\Lambda^2+2}\,e_3, \Lambda\sqrt{2}\,(e_1\pm e_2 \pm e_3).
\end{multline}
In the second case the system $F_3^2(\Lambda)$ consists of the
covectors
\begin{multline*}
\sqrt{2\Lambda^2+1}\,(e_1 \pm e_2), \sqrt{2}\,(e_2 \pm e_3), \sqrt{2}\,(e_1 \pm e_3), \\ 2\sqrt{2\Lambda^2+1}\,e_3, 2\Lambda e_1, 2\Lambda e_2, \Lambda(e_1\pm e_2 \pm 2 e_3).
\end{multline*}
Both systems contain 13 covectors. It turns out that these two families are
equivalent as the families of $\vee$-systems. More precisely, the
following statement holds.
\begin{prop}
The $\vee$-system $F_3^1(\lambda)$ is equivalent to the
$\vee$-system $F_3^2(\mu)$ with $ \mu=\frac{1}{2\lambda}$.
\end{prop}
Indeed, a linear transformation $\cal A,$ which transforms $F_3^1(\lambda)$
to $F_3^2(\mu),$ has the form
$${\cal A}(\sqrt{4\lambda^2+2}\, e_1)=\sqrt{2\mu^2+1} \,(e_1+e_2),
{\cal A}(\sqrt{4\lambda^2+2} \,e_2)=\sqrt{2\mu^2+1}\, (e_1-e_2), $$ $$
{\cal
A}(\sqrt{4\lambda^2+2}\, e_3)=2\sqrt{2\mu^2+1} \,e_3.
$$
This equivalence is related to the symmetry
$ \Lambda \rightarrow (2\Lambda)^{-1}$
of the initial $F_4$-system (\ref{F4}), which can be easily checked.

We will denote the family $F_3^1(\Lambda)$ simply as $F_3(\Lambda)$.
As it follows from the proposition the parameter $\Lambda$ here is natural to be considered as a point on a projective line ${\mathbb R}P^1$ rather than in $\R.$

We would like to mention that the one-parameter families $F_4$,
$F_3$ cannot be generalised to two-parametric families of
$\vee$-systems. More precisely, a simple calculation shows that if
we put two non-trivial arbitrary coefficients at the vectors of type
$e_i$ and at $e_1\pm e_2 \pm e_3 \pm e_4$ in \mref{F4}  we will have
a $\vee$-system only in the case \mref{F4} ( similarly for
\mref{f41}).

This calculation suggests to consider the following systems,
which we will call the systems of $F_n$-type:
\beq{gener}
e_i \pm e_j, \Lambda e_i, M(e_1\pm e_2 \pm \ldots \pm e_n),
\eeq
where $M$ is assumed to be non-zero.

\begin{theorem}
In dimension $n=5$ the set \mref{gener}  is a $\vee$-system if and
only if $M^2=1$ and $\Lambda^2 = 6$.  In dimension $n=6$  this is
true if and only if $\Lambda^2 =4$ and $M^2=\frac{1}{2}.$  There are
no $\vee$-systems of $F_n$-type for $n>6.$
\end{theorem}

{\bf Proof.} The system is invariant under the action of the Coxeter group
$B_n,$ therefore the corresponding $\vee$-inner product must be proportional
to the standard Euclidean one. The $\vee$-conditions are non-trivial only for planes, containing 2 covectors of the form $M(e_1\pm e_2 \pm \ldots \pm e_n).$

If $n=5$ there are only two different cases. For the plane, containing the covectors
$$
M(e_1+e_2+e_3+e_4+e_5), M(-e_1+e_2+e_3+e_4+e_5), \Lambda e_1
$$
the $\vee$-condition gives $\Lambda^2=6 M^2$.
For the plane containing $$
M(e_1+e_2+e_3+e_4+e_5), M(-e_1-e_2+e_3+e_4+e_5), e_1+e_2
$$
the $\vee$-condition is $M^2=1$.

Similarly in dimension 6 we have two types of planes with non-trivial $\vee$-conditions:
those containing the covectors
$$
M(e_1+e_2+e_3+e_4+e_5+e_6), M(-e_1+e_2+e_3+e_4+e_5+e_6), \Lambda e_1
$$
and
$$
M(e_1+e_2+e_3+e_4+e_5+e_6), M(-e_1-e_2+e_3+e_4+e_5+e_6), e_1+e_2.
$$
The corresponding $\vee$-conditions are  $8M^2=\Lambda^2$ and $2M^2=1$
respectively.

One can check that the $\vee$-conditions for the systems (\ref{gener}) with $n>6$
leads to $M=0.$  For example, for $n=7$ the covectors $M (e_1 + e_2 + e_3 + e_4 +  e_5 + e_6 + e_7)$ and $M (e_1 + e_2 + e_3 - e_4 -  e_5 - e_6 - e_7)$  are the only covectors from the corresponding plane and therefore must be orthogonal, which is the case only if $M=0.$ The theorem is proved.

We denote the corresponding $\vee$-systems in dimension 5 and 6 as
$F_5$ and $F_6$ respectively. The system $F_6$ contains 68
covectors. Its restriction along $e_6$ gives the system $F_5,$
containing 41 covector. The restriction of $F_5$ along $e_5$ gives a
system from the $F_4$ family.

{\bf Remark.} In $\R^8$ there is a possibility to have non-zero $M$
by considering \mref{gener} with the additional requirement that the
numbers of negative signs in the covectors
$$
M(e_1\pm e_2\pm \ldots \pm e_8)
$$
are even. Then the choice $M=1/2, \Lambda=0$ does lead to a
$\vee$-system, but this is simply the (positive part of) root system
$E_8$.

It turns out that the system $F_6$ itself is a restriction of the
root system $E_8$ along its subsystem $e_7 \pm e_8$ as we will see in the next section.

\subsection{Other exceptional systems}

To list all the $\vee$-systems coming from the exceptional root systems we need the results of Orlik and Solomon \cite{OS} and Shcherbak \cite{Shch}, who classified the strata in the Coxeter discriminants.
We will use the standard terminology of the theory of Coxeter groups, for which we refer to \cite{Hum}.

Let $G$ be a finite group generated by reflections in a real vector space $V.$
The reflection hyperplanes divide $V$ into several connected components called {\it chambers.}
Choose a chamber $C$ and consider the set $S$  of the reflections in the hyperplanes, which bound $C$.  The set $S$ can be identified with the vertices of the corresponding {\it Coxeter graph} $\Gamma$ (see \cite{Hum}). For any subset $J \subset S$
the corresponding {\it parabolic subgroup} $G_J$ is generated by the reflections from $J.$

There exists the following natural correspondence between the strata and the conjugacy classes of the parabolic subgroups of the corresponding Coxeter group.
Let  $L$ be a linear subspace of $V$, which is an intersection of some reflection hyperplanes.
If $R$ is the Coxeter root system of $G$ then $L = L_B$ for some subsystem $B \subset R$ (see section 4). Consider the subgroup $H$ generated by reflections $s_\b, \b\in B$, which leaves the subspace $L$ fixed. It is known that $H$ is conjugated to $G_J$ for some $J \subset S.$ This gives us a one-to-one correspondence between the orbits of $G$ on the set of all such $L$ (or equivalently, the strata in the Coxeter discriminant) and the parabolic subgroups of $G$ considered up to a conjugation.

The explicit description of the corresponding subgraphs (with the set of vertices $J$) of the Coxeter graphs for the exceptional groups can be found in \cite{OS, Shch}. Typically all the subgroups of a given type are conjugate besides the following exceptions listed below.

As $\vee$-systems are non-trivial only in dimension greater than two
we are interested in the parabolic subgroups of corank at least 3.
Then there are only two pairs of non-conjugate isomorphic subgroups
in the group $E_7$, and one more pair in the group $F_4$.

\setlength\unitlength{0.8cm}

\newcommand{\Esevena}{
\begin{picture}(8,1.5)(0,1.8)
\put(1,2){\circle{0.2}} \put(1.1,2){\line(1,0){.8}}
\put(2,2){\circle*{0.2}} \put(2.1,2){\line(1,0){.8}}
\put(3,2){\circle{0.2}} \put(3.1,2){\line(1,0){.8}}
\put(4,2){\circle*{0.2}} \put(4.1,2){\line(1,0){.8}} \put(5
,2){\circle{0.2}} \put(3  ,2.1){\line(0,1){.8}} \put(3
,3){\circle{0.2}} \put(5.1,2){\line(1,0){.8}}
\put(6,2){\circle*{0.2}}
\end{picture}
}
\newcommand{\Esevenb}{
\begin{picture}(8,1.5)(0,1.8)
\put(1,2){\circle{0.2}} \put(1.1,2){\line(1,0){.8}}
\put(2,2){\circle{0.2}} \put(2.1,2){\line(1,0){.8}}
\put(3,2){\circle{0.2}} \put(3.1,2){\line(1,0){.8}}
\put(4,2){\circle*{0.2}} \put(4.1,2){\line(1,0){.8}} \put(5
,2){\circle{0.2}} \put(3  ,2.1){\line(0,1){.8}} \put(3
,3){\circle*{0.2}} \put(5.1,2){\line(1,0){.8}}
\put(6,2){\circle*{0.2}}
\end{picture}
}
\newcommand{\Esevenc}{
\begin{picture}(8,1.5)(0,1.8)
\put(1,2){\circle{0.2}} \put(1.1,2){\line(1,0){.8}}
\put(2,2){\circle{0.2}} \put(2.1,2){\line(1,0){.8}}
\put(3,2){\circle{0.2}} \put(3.1,2){\line(1,0){.8}}
\put(4,2){\circle*{0.2}} \put(4.1,2){\line(1,0){.8}} \put(5
,2){\circle*{0.2}} \put(3  ,2.1){\line(0,1){.8}} \put(3
,3){\circle*{0.2}} \put(5.1,2){\line(1,0){.8}}
\put(6,2){\circle*{0.2}}
\end{picture}
}
\newcommand{\Esevend}{
\begin{picture}(8,1.5)(0,1.8)
\put(1,2){\circle{0.2}} \put(1.1,2){\line(1,0){.8}}
\put(2,2){\circle*{0.2}} \put(2.1,2){\line(1,0){.8}}
\put(3,2){\circle{0.2}} \put(3.1,2){\line(1,0){.8}}
\put(4,2){\circle*{0.2}} \put(4.1,2){\line(1,0){.8}} \put(5
,2){\circle*{0.2}} \put(3  ,2.1){\line(0,1){.8}} \put(3
,3){\circle{0.2}} \put(5.1,2){\line(1,0){.8}}
\put(6,2){\circle*{0.2}}
\end{picture}
}

Namely, in the case of the group $E_7$ there are two non-conjugate classes
of subgroups of type $A_1^3$. The Coxeter subgraphs corresponding to
them can be chosen as follows:

$(E_7, A_1^3)_1$: $\Esevenb$

$(E_7, A_1^3)_2$: $\Esevena$

\noindent Also there are two non-conjugate classes of subgroups of
type $A_1 \times A_3$ inside $E_7$. The corresponding graphs can be
taken as follows:

$(E_7, A_1 \times A_3)_1$: $\Esevenc$

$(E_7, A_1 \times A_3)_2$: $\Esevend$

\noindent Finally, the system $F_4$ has 2 types of non-conjugate
subgraphs of type $A_1$ given by the two roots of different length.

Applying the restriction procedure from the section 3 we can construct a $\vee$-system
for each pair $(G,H),$ where $G$ is a Coxeter group and $H$ is its parabolic subgroup.
The complete list of the corresponding $\vee$-systems for the exceptional Coxeter groups is given in the Appendix.

The $\vee$-systems $F_3, F_5, F_6$ found in the section
\ref{exceptF}  correspond to the following $\vee$-systems from the
Appendix: $$ F_6 = (E_8, A_1 \times A_1), \,\, F_5= (E_8,
A_3),\,\, F_3(\Lambda) = (F_4(\Lambda), A_1)_1. $$ In particular,
we see that $F_5$ and $F_6$ are indeed the restrictions of
$E_8$-system.

\section{Non-Coxeter families of $ \pmb \vee$-systems}\label{noncoxeter}

A natural question is if there exist $\vee$-systems, which can not
be obtained through the restriction of the Coxeter root systems. In
this section we are going to show that the answer is positive by
presenting some new one-parameter families of $\vee$-systems, which
contain only a finite number of restrictions of Coxeter root
systems. Note that the restrictions of the exceptional Coxeter root
systems (unlike the classical ones) form a set, which is already
closed in Zariski sense.

\begin{theorem}\label{th5}
The following set of covectors in $\R^4$ \beq{oneparam} e_1 \pm e_2,
e_1 \pm e_3, e_2 \pm e_3, \Lambda e_1, \Lambda e_2, \Lambda e_3, K
e_4, M(e_1\pm e_2 \pm e_3 \pm e_4) \eeq with $M \neq 0$ is a
$\vee$-system if and only if the parameters satisfy \beq{param}
\Lambda^2=2(2M^2+1), \quad K^2 = \frac{2M^2(2M^2-1)}{M^2+1}. \eeq
The corresponding $\vee$-system is a restriction of a Coxeter root
system if and only if $M^2= 1$ or $M^2= \frac{1}{2}.$
\end{theorem}
{\bf Proof.} The $\vee$-quadratic form in this case is
$$
G=(\Lambda^2+8M^2+4)(x_1^2+x_2^2+x_3^2)+(K^2+8 M^2)x_4^2.
$$
The $\vee$-conditions are non-trivial for the following three types
of the two-dimensional planes
$$
\pi_1= <M(e_1+e_2+e_3+e_4), \Lambda e_1>, G_1=(2M^2 +\Lambda^2)x_1^2
+ 2M^2 (x_2+x_3+x_4)^2,
$$
$$
\pi_2= <M(e_1+e_2+e_3+e_4), Ke_4>, G_2= 2M^2 (x_1+x_2+x_3)^2+ (2M^2
+K^2)x_4^2,
$$
$$
\pi_3= <M(e_1+e_2+e_3+e_4), e_1+e_2>, G_3=(2M^2 + 1)(x_1+x_2)^2 +
2M^2 (x_3+x_4)^2.
$$
and have the form
$$
\Lambda^2+2M^2 =
2M^2(\Lambda^2+8M^2+4)(\frac{2}{\Lambda^2+8M^2+4}+\frac1{K^2+8M^2}),
$$
$$
(\Lambda^2+8M^2+4)(2M^2+K^2)=6M^2(K^2+8M^2),
$$
$$
2M^2+1=M^2(1+\frac{\Lambda^2+8M^2+4}{K^2+8M^2}).
$$
One can check that these relations are equivalent to the parametrisation \mref{param}.

When $M^2= 1$ or $M^2= \frac{1}{2}$ these $\vee$-systems
are equivalent to the Coxeter restrictions $(E_7,A_3)$ and $(E_6, A_1
\times A_1)$ respectively.  Note that these are the only 4-dimensional Coxeter restrictions,
containing 18 and 17 covectors respectively. One can check that other values of $M$ do not correspond to any of these two systems. This completes the proof of the theorem.

More examples of the non-Coxeter families one can get by the restrictions of the $\vee$-systems \mref{oneparam}, \mref{param}
to the corresponding 3-dimensional hyperplanes determined by one of the covectors.
In particular, for the covector $e_i \pm e_j$ we have the
one-parameter family of $\vee$-systems
$$
\sqrt{2(2M^2+1)} e_1, \, 2 \sqrt{2(M^2+1)} e_2, M
\sqrt{\frac{2(2M^2-1)}{M^2+1}}e_3, \, $$ $$\sqrt{2}(e_1\pm e_2), \,
M\sqrt{2}(e_1 \pm e_3), \, M(e_1 \pm 2 e_2 \pm e_3).
$$
This family contains (for non-zero $M$) only two Coxeter restrictions: $(E_7, A_1 \times A_3)_2$ and  $(E_6, A_1^3)$
when $M^2= 1$ and $M^2= \frac{1}{2}$ respectively.

For the covector $e_1+e_2+e_3-e_4$ we have the
following family of $\vee$-systems
$$
e_1+e_2, e_1+e_3, e_2+e_3, \sqrt{2}e_1, \sqrt{2}e_2, \sqrt{2}e_3,
\frac{M\sqrt{2}}{\sqrt{M^2+1}}(e_1+e_2+e_3),
$$
$$
\frac1{\sqrt{4M^2+1}}(e_1-e_2), \frac1{\sqrt{4M^2+1}}(e_1-e_3),
\frac1{\sqrt{4M^2+1}}(e_2-e_3).
$$
This $\vee$-system is equivalent to Coxeter restriction $(E_7, A_4)$
when  $M^2= 1$ and to $(E_6, A_1 \times A_2)$ at
$M^2= \frac{1}{2}$, for other $M \neq 0$ it can not be obtained in such a way.

Note that when $M=0$ all these systems reduce to some Coxeter root
systems.

{\bf Remark.} The $\vee$-systems in Theorem 4 are equivalent to the
deformed root systems related to Lie superalgebra of type $AB(1,3)$,
which were described by Sergeev and Veselov \cite{SV1}. One can show
that other exceptional simple Lie superalgebras also give
non-Coxeter families of $\vee$-systems.



\section{Concluding remarks}

The restrictions of the Coxeter root systems on the mirror
hyperplanes and more generally to any their intersections were
investigated within the general theory of the hyperplanes
arrangements by Orlik and Solomon \cite{OS}. We have shown that the
corresponding complements admit a natural logarithmic Frobenius
structure, dual in Dubrovin's sense to the Strachan's structures on
the strata of the Coxeter discriminants.

As we know the general logarithmic Frobenius structures are related
to the $\vee$-systems, which can be considered as a proper extension
of the Coxeter root systems. We have shown that this class of
systems is closed under the operation of restriction to any
intersection of the corresponding hyperplanes. The restrictions of
the Coxeter root systems give us many examples of the $\vee$-systems
but as we have shown not all of them.

Thus the classification of the $\vee$-systems still remains  one of
the most important open problems in this area. Probably the next
step in this direction should be classification of all families of
$\vee$-systems passing through these restrictions. It would also be
natural to study $\vee$-systems in the complex space, in particular,
in relation with complex reflection groups.

Another very important problem is to investigate the relations with
Seiberg-Witten theory, see \cite{MMM, D2}. In particular, it would
be interesting to analyse from this point of view a role of the
special $\vee$-systems related to the deformed Calogero-Moser
systems and Lie superalgebras \cite{SV1}. The relations with the
theory of (super)Jack polynomials (see \cite{SV2, KMSV}) also
deserve better understanding.

\newpage

\vspace{5mm} \noindent {\bf Acknowledgements.}

\medskip

We are grateful to the organisers of the programme on Combinatorial Aspects of Integrable Systems  (RIMS, Kyoto) in July 2004, where this work was mainly done. One of us (M.F.) would like to thank also Prof. T. Miwa and S. Loktev for the possibility to visit RIMS during that time.

We would like to acknowledge the useful and stimulating discussions
we had with O.A. Chalykh, B.A. Dubrovin, C. Hertling, G. Lehrer,
V.P. Leksin, D. Panov, A.N. Sergeev and I. Strachan.

This work was partially supported by the European research network ENIGMA (contract MRTN-CT-2004-5652) and ESF programme MISGAM. M.F. also
acknowledges the support of Chapman Fellowship at the Mathematics
Department of Imperial College London.

\vspace{1cm}

\appendix

\begin{center}

{\Large Appendix: \bf Restrictions of the exceptional Coxeter root
systems}

\end{center}

Below is a complete list of the $\vee$-systems $A$, which can be obtained as the restrictions of the exceptional Coxeter root systems. They are labeled by a pair $(G,H)$, where $G$ is an exceptional Coxeter group and $H$ is its parabolic subgroup (see section 5). When the type of the subgroup does not fix the subgroup up to a conjugation we use the index 1 or 2 following the description of all such cases given above. We give also the dimension of the space spanned by the $\vee$-system $A$ (which is the same as corank of $H$) and the number $|A|$ of (noncollinear) covectors in $A.$
The list of the equivalences between these $\vee$-systems is given after the table.



\begin{longtable}{|c|c|p{83mm}|c|p{6mm}|}

\hline   & $(G,H)$   &  Covectors of the $ \vee$-system $A$ &
Dimension & $|A|$  \hspace{2cm}  \\

\hline 1 &  $(E_8, A_1)$ & $e_i \pm e_j (1\le i<j\le 6)$;
$\sqrt{2}(e_i \pm e_7) (1\le i\le 6)$; $2 e_7$;
$\frac{\sqrt{2}}{2} (e_1\pm e_2 \pm e_3 \pm e_4 \pm e_5 \pm e_6)$
(odd number of minuses); $\frac12 (e_1\pm e_2 \pm e_3 \pm e_4 \pm
e_5 \pm e_6 \pm 2e_7)$ (even number of minuses in the first six
terms)  & 7 & 91\\

\hline 2 & $(E_8,A_1 \times A_1)$ & $ e_i \pm e_j (1\le i<j\le
6);$ $2 e_i  (1\le i \le 6)$;  $ \frac{\sqrt{2}}{2} (e_1\pm e_2
\pm e_3 \pm e_4 \pm e_5 \pm e_6)$ & 6 & 68 \\

\hline 3 & $(E_8,A_2)$ & $ e_i \pm e_j (1\le i<j\le 5)$;
$\sqrt{3}(e_i \pm e_6) (1\le i \le 5)$;  $2\sqrt{3} e_6$;
$\frac{\sqrt{3}}{2} (e_1\pm e_2 \pm e_3 \pm e_4 \pm e_5 \pm e_6)$
(odd number of minuses); $\frac12 (e_1\pm e_2 \pm e_3 \pm e_4 \pm
e_5 \pm 3 e_6)$ (even number of minuses) & 6  & 63 \\

\hline 4 & $(E_8,A_1^3)$ & $ e_i \pm e_j \,\, (1\le i<j\le 4)$;
$\sqrt{2}(e_i \pm e_5), 2 e_i (1\le i \le 4)$; $2\sqrt{3} e_5$;  $
e_1\pm e_2 \pm e_3 \pm e_4$; $\frac{\sqrt{2}}{2} (e_1\pm e_2 \pm
e_3 \pm e_4 \pm 2 e_5) $ & 5 & 49  \\

\hline 5 & $(E_8,A_1 \times A_2 )$ & $ e_i \pm e_j \,\,(1\le
i<j\le 3)$; $\sqrt{2}(e_i \pm e_4), \sqrt{3}(e_i \pm e_5)\,(1\le i
\le 3)$; $\sqrt{6}(e_4 \pm e_5)$;  $2 e_4$; $2\sqrt{3} e_5$;
$\frac{\sqrt{3}}{2} (e_1\pm e_2 \pm e_3 \pm e_5 \pm 2 e_4 )$ \,
(odd number of minuses in the first four terms);  $\frac12 (e_1\pm
e_2 \pm e_3  \pm 3 e_5 \pm 2 e_4)$ (even number of minuses in the
first four terms); $\frac{\sqrt{6}}{2} (e_1\pm e_2 \pm e_3 \pm
e_5)$  (even number of minuses);  $\frac{\sqrt{2}}{2} (e_1\pm e_2
\pm e_3 \pm 3 e_5)$ (odd number of minuses) & 5 &  46 \\

\hline 6 & $(E_8,A_3)$ & $e_i \pm e_j \,  (1\le i<j\le 5)$;
$\sqrt{6} e_i (1\le i \le 5)$;  $ e_1\pm e_2 \pm e_3 \pm e_4 \pm
e_5 $ & 5 & 41 \\

\hline 7 & $(E_8,A_1^4)$ & $ \sqrt{3}(e_1 \pm e_2)$, $\sqrt{2}(e_1
\pm e_3)$, $\sqrt{2}(e_2 \pm e_3)$, $\sqrt{2}(e_1 \pm e_4)$,
$\sqrt{2}(e_2 \pm e_4)$, $2(e_3 \pm e_4)$;  $ 2 e_1, 2 e_2, 2
\sqrt{3} e_3, 2 \sqrt{3} e_4$; $ e_1\pm e_2 \pm 2 e_3$; $e_1\pm
e_2 \pm 2 e_4; \, \frac{\sqrt{2}}{2}(e_1\pm e_2 \pm 2 e_3 \pm 2
e_4) $ & 4 &  32\\

\hline 8 & $(E_8,A_1^2 \times A_2)$ & $ e_i \pm e_j \,(1\le i<j
\le 3)$; $\sqrt{3}(e_i \pm e_4) \,(1\le i \le 3)$; $ 2 e_1, 2 e_2,
2 e_3, 2 \sqrt{6} e_4; $ $ \frac{\sqrt{6}}{2}(e_1\pm e_2 \pm e_3
\pm e_4)$; $ \frac{\sqrt{2}}{2}(e_1\pm e_2 \pm e_3 \pm 3e_4)$ & 4
& 32
\\

\hline 9 & $(E_8,A_2^2)$ & $e_1 \pm e_2$,  $\sqrt{3}(e_1 \pm
e_3)$, $\sqrt{3}(e_1 \pm e_4)$, $\sqrt{3}(e_2 \pm e_3)$,
$\sqrt{3}(e_2 \pm e_4)$, $3(e_3 \pm e_4)$;  $2 \sqrt{3} e_3, 2
\sqrt{3} e_4$; $\frac12(e_1\pm e_2 \pm 3 e_3 \pm 3 e_4)$ (even
number of minuses); $\frac{\sqrt{3}}{2}(e_1\pm e_2 \pm 3 e_3 \pm
e_4)$ (odd number of minuses); $\frac{\sqrt{3}}{2}(e_1\pm e_2 \pm
e_3 \pm 3 e_4)$ (odd number of minuses); $\frac32(e_1\pm e_2 \pm
e_3 \pm e_4)$ (even number of minuses) & 4 & 30 \\

\hline 10 & $(E_8, A_1 \times A_3)$ & $e_i \pm e_j \,(1\le i<j \le
3)$; $\sqrt{2}(e_i \pm e_4) \,(1\le i \le 3)$;
  $ \sqrt{6} e_1, \sqrt{6} e_2,
\sqrt{6} e_3,  4 e_4$; $ e_1\pm e_2 \pm e_3 \pm 2 e_4$;
$\sqrt{2}(e_1\pm e_2 \pm e_3)$ & 4 &  28\\

\hline 11 & $(E_8,A_4)$ & $ e_i \pm e_j \,(1\le i<j \le 3);
\sqrt{5}(e_i \pm e_4) \,\,(1\le i \le 3)$;
 $2 \sqrt{10} e_4$;  $\frac12(e_1\pm e_2 \pm
e_3 \pm 5 e_4)$ (even number of minuses); $\frac{\sqrt{5}}2(e_1\pm
e_2 \pm e_3 \pm 3 e_4)$ (odd number of minuses);
$\frac{\sqrt{10}}2(e_1\pm e_2 \pm e_3 \pm e_4)$ (even number of
minuses) & 4 & 25 \\

\hline 12 & $(E_8,D_4)$ & $ e_i \pm e_j (1\le i<j \le 4)$; $2
\sqrt{2} e_i (1\le i  \le 4)$; $\sqrt{2}(e_1\pm e_2 \pm e_3 \pm
e_4)$ & 4 & 24\\

\hline 13 & $(E_8,A_1^3 \times A_2)$ & $ \sqrt{2}(e_1 \pm e_2)$,
$\sqrt{6}(e_2 \pm e_3)$, $\sqrt{6}(e_1 \pm e_3)$, $2e_1$,
$2\sqrt{3} e_2$, $2 \sqrt{6} e_3$,  $\frac{\sqrt{2}}2(e_1\pm 2 e_2
\pm 3 e_3)$, $\frac{\sqrt{6}}{2}(e_1\pm 2 e_2 \pm  e_3)$, $e_1\pm
3 e_3$ & 3 & 19 \\

\hline 14 & $(E_8,A_2^2 \times A_1)$ & $ \sqrt{3}(e_1 + e_2)$, $
3(e_1 + e_3)$, $3(e_2 + e_3)$, $e_1- e_2$, $\sqrt{3}(e_1-e_3)$,
$\sqrt{3}(e_2-e_3)$, $\sqrt{6} e_1$, $\sqrt{6} e_2$, $6 \sqrt{2}
e_3$,  $\sqrt{6}(e_1 + e_2 + 3 e_3)$, $\sqrt{3}(e_1 + e_2 + 4
e_3)$, $3 (e_1 + e_2 + 2 e_3)$, $e_1 + 2 e_2 + 3 e_3$,
$\sqrt{3}(e_1 + 3 e_3)$,  $\sqrt{3}(e_2 + 3 e_3)$, \, $2 e_1 + e_2
+ 3e_3$, \, $\sqrt{6}(e_1 +2 e_3)$, $\sqrt{6}(e_2 + 2 e_3)$,
$\sqrt{6}(e_1 + e_2 +e_3)$ &  3 & 19 \\

\hline 15 & $(E_8,A_1^2 \times A_3)$ & $ 2(e_1 \pm e_2)$, $2(e_2
\pm e_3)$, $2(e_1 \pm e_3)$, $2e_1$, $2\sqrt{10} e_2$, $2 e_3$, $
\frac{\sqrt{2}}2(e_1\pm 4 e_2 \pm e_3)$, $\sqrt{2}(e_1\pm 2 e_2
\pm e_3)$ & 3 & 17 \\

\hline 16 & $(E_8,A_2 \times A_3)$ & $ 2\sqrt{3}(e_1 \pm e_2)$,
$2(e_2 \pm e_3)$, $\sqrt{3}(e_1 + e_3)$, $\sqrt{\frac{15}{2}}(e_1-
e_3)$, $2\sqrt{3}e_1$, $2\sqrt{6}e_2$, $\frac12(e_3 + 3 e_1 \pm 4
e_2)$, $ e_3 - 3 e_1 \pm 2 e_2$, $\frac{\sqrt{6}}2 (e_3 + 3 e_1)$,
$\frac{\sqrt{3}}2 (e_3 - e_1 \pm 4 e_2)$, $\sqrt{3} (e_3 + e_1 \pm
2 e_2)$ & 3 & 17 \\

\hline 17 & $(E_8,A_1 \times A_4)$ & $\sqrt{10}(e_1 \pm e_2)$,
$\sqrt{2}(e_1 \pm e_3)$, $\sqrt{5}(e_2 + e_3)$, $\sqrt{10}(e_2 -
e_3)$, $2e_1$, $2\sqrt{10} e_2$,  $\frac12(e_3 \pm 2 e_1 + 5
e_2)$, $\frac{\sqrt{5}}2(e_3\pm 2 e_1 - 3 e_2)$,
$\frac{\sqrt{10}}2(e_3\pm 2 e_1 + e_2)$, $\frac{\sqrt{2}}2(e_3 - 5
e_2)$, $\frac{\sqrt{10}}2(e_3 + 3 e_2)$ &  3 & 16 \\

\hline 18 & $(E_8,A_1 \times D_4)$ & $ \sqrt{2}(e_1 \pm e_2)$,
$\sqrt{2}(e_1 \pm e_3)$, $\sqrt{5}(e_2 \pm e_3)$, $2\sqrt{5}e_1$,
$ 2\sqrt{2} e_2$, $2\sqrt{2} e_3$, $\sqrt{2}(e_2 \pm e_3 \pm 2
e_1)$. & 3 & 13 \\

\hline 19 & $(E_8,A_5)$ & $ \sqrt{6}(e_1 \pm e_2)$, $\sqrt{6}(e_2
\pm e_3)$, $\sqrt{6}(e_1- e_3)$, $e_1 + e_3$, $ 2\sqrt{15}e_2$, $
\frac12(e_1 \pm 6 e_2 +  e_3)$, $\frac{\sqrt{6}}2(e_1\pm 4 e_2 -
e_3)$, $\frac{\sqrt{15}}2(e_1 \pm 2 e_2 + e_3)$ &  3 & 13 \\

\hline 20 & $(E_8,D_5)$ & $e_1 \pm e_2$, $e_2 \pm e_3$, $e_1 \pm
e_3$, $\sqrt{10}e_1$,  $\sqrt{10} e_2$, $\sqrt{10} e_3$,  $2(e_1
\pm e_2 \pm e_3)$ & 3 & 13 \\ \hline


\hline

\hline 21 & $(E_7,A_1)$ & $ e_i \pm e_j$ $(1\le i<j\le 4)$;
$\sqrt{2}(e_5 \pm e_i)$ $(1\le i \le 4)$; $2 e_5$, $e_6$, $
\frac{\sqrt{2}}{2} (e_6 \pm e_1 \pm e_2 \pm e_3 \pm e_4)$ (even
number of minuses);  $ \frac12 (e_6 \pm 2 e_5 \pm e_1 \pm e_2 \pm
e_3 \pm e_4)$ (odd number of minuses in the last four terms) &  6
& 46 \\

\hline 22 & $(E_7,A_1\times A_1)$ & $ e_i \pm e_j$ $(1\le i<j\le
4)$; $2e_1$, $2e_2$, $2e_3$, $2e_4$, $e_5$,   $
\frac{\sqrt{2}}{2}(e_5 \pm e_1 \pm e_2 \pm e_3 \pm e_4)$ &  5 & 33
\\

\hline 23 & $(E_7,A_2)$ & $ e_1 \pm e_2,  e_1 \pm e_3, e_2 \pm
e_3$, $\sqrt{3}(e_4\pm e_1)$, $\sqrt{3}(e_4\pm e_2)$,
$\sqrt{3}(e_4\pm e_3)$, $2\sqrt{3}e_4$, $e_5$, $\frac{\sqrt{3}}{2}
(e_5 \pm e_4 \pm e_1 \pm e_2 \pm e_3)$ (even number of minuses);
$ \frac12 (e_5 \pm 3 e_4 \pm e_1 \pm e_2 \pm e_3)$ (odd number of
minuses) &  5 & 30 \\

\hline 24 & $(E_7,A_1^3)_1$ & $2(e_i \pm e_j)$ $(1 \le i <j \le
4)$, $2e_i$ $(1 \le i \le 4)$, $e_1 \pm e_2 \pm e_3 \pm e_4$ &  4
& 24
\\

\hline 25 & $(E_7,A_1^3)_2$ & $2(e_1\pm e_2)$, $2(e_1\pm e_3)$,
$2(e_2\pm e_3)$, $2e_1$, $2e_2$, $2e_3$, $2\sqrt{3}e_4$,  $
\sqrt{2}(e_4 \pm e_1 \pm e_2)$, $\sqrt{2}(e_4\pm e_1 \pm e_3)$, $
\sqrt{2}(e_4 \pm e_2 \pm e_3)$ &  4 & 22 \\

\hline 26 & $(E_7,A_1 \times A_2)$ & $ \sqrt{2}(e_2\pm e_3)$,
$\sqrt{3}(e_3 \pm e_1)$, $\sqrt{6}(e_2 \pm e_1)$, $2e_2$,
$2\sqrt{3}e_1$, $e_4$, $ \frac12(e_4 \pm 2 e_2 - e_3 + 3 e_1)$,
$\frac12(e_4 \pm 2 e_2 + e_3 - 3 e_1)$, $ \frac{\sqrt{3}}{2}(e_4
\pm 2 e_2 + e_3 + e_1)$, $\frac{\sqrt{3}}{2}(e_4 \pm 2 e_2 - e_3 -
e_1)$, $ \frac{\sqrt{2}}{2}(e_4 + e_3 + 3 e_1)$,
$\frac{\sqrt{2}}{2}(e_4 - e_3 - 3 e_1)$, $ \frac{\sqrt{6}}{2}(e_4
+ e_3 - e_1)$, $ \frac{\sqrt{6}}{2}(e_4 - e_3 + e_1) $ &  4 & 21
\\

\hline 27 & $(E_7,A_3)$ & $ e_1\pm e_2$, $ e_1\pm e_3$, $ e_2\pm
e_3$, $\sqrt{6}e_1$, $\sqrt{6}e_2$, $\sqrt{6}e_3$, $e_4$, $e_4 \pm
e_1 \pm e_2 \pm e_3$ &  4 & 18 \\

\hline 28 & $(E_7,A_1^4)$ & $2(e_1\pm e_2)$, $ 2(e_1\pm e_3)$, $
2(e_2\pm e_3)$, $2\sqrt{3}e_1$, $2\sqrt{3}e_2$,  $2\sqrt{3}e_3$,
$\sqrt{2}(e_1 \pm e_2 \pm  e_3)$ &  3 & 13 \\

\hline 29 & $(E_7,A_1^2 \times A_2)$ & $ \sqrt{3}(e_1\pm e_2)$,
$2e_1$, $2\sqrt{6}e_2$, $e_3$,  $ \frac{\sqrt{2}}2 (e_3 \pm e_1
\pm 3 e_2)$, $\frac{\sqrt{6}}2( e_3 \pm e_1 \pm e_2) $ &  3 & 13
\\

\hline 30 & $(E_7,A_2^2)$ & $ 3(e_2\pm e_1)$, $2 \sqrt{3}e_2$, $2
\sqrt{3}e_1$, $e_3$, $\frac12 (e_3 + 3 e_2 - 3 e_1)$, $\frac12(
e_3 -3 e_2 + 3 e_1)$, $ \frac{\sqrt{3}}2 (e_3 +3 e_2+ e_1)$,
$\frac{\sqrt{3}}2 (e_3 -3e_2- e_1)$, $\frac{\sqrt{3}}2 (e_3 +
e_2+3 e_1)$, $ \frac{\sqrt{3}}2 (e_3 -e_2- 3 e_1)$, $\frac32 (e_3
+ e_2 -e_1)$, $\frac32 (e_3 - e_2+e_1)$ &  3 & 13 \\

\hline 31 & $(E_7,A_1 \times A_3)_1$ & $ 2\sqrt{2}(e_1\pm e_3)$,
$2e_1$, $2\sqrt{6}e_3$, $2e_2$,  $e_2 \pm  e_1 \pm 2 e_3$,
$\sqrt{6}(e_2 \pm e_1)$, $2\sqrt{2}(e_2 \pm e_3)$ &  3 & 13 \\

\hline 32 & $(E_7,A_1 \times A_3)_2$ & $ 2\sqrt{2}(e_1\pm e_3)$,
$2e_1$,\, $2\sqrt{6}e_3$, $2e_2$,  $ e_2 \pm 2 e_1 \pm 2 e_3$,
$\frac{\sqrt{2}}2(e_2 \pm 4 e_3)$ &  3 & 11 \\

\hline 33 & $(E_7,A_4)$ & $ \sqrt{5}(e_1\pm e_2)$, $2
\sqrt{10}e_2$, $e_3$, $\frac12 (e_3 - e_1 + 5 e_2)$, $\frac12( e_3
+ e_1 - 5 e_2)$, $ \frac{\sqrt{5}}2 (e_3 +e_1+3 e_2)$,
$\frac{\sqrt{5}}2 (e_3 -e_1-3 e_2)$, $\frac{\sqrt{10}}2 (e_3 -e_1+
e_2)$, $\frac{\sqrt{10}}2 (e_3 +e_1- e_2)$ &  3 & 10 \\

\hline 34 & $(E_7,D_4)$ & $ e_1\pm e_2$,  $2\sqrt{2}e_1$,
$2\sqrt{2}e_2$, $e_3$, $\sqrt{2} (e_3 \pm e_1 \pm e_2)$ &  3 & 9
\\ \hline


\hline

\hline 35 & $(E_6,A_1)$ & $ e_1 \pm e_2$, $ e_1 \pm e_3$, $ e_2
\pm e_3$, $\sqrt{2}(e_4 \pm e_1)$, $\sqrt{2}(e_4 \pm e_2)$,
$\sqrt{2}(e_4 \pm e_3)$, $2 e_4$, $ \frac{\sqrt{2}}{2} (e_5 \pm
e_1 \pm e_2 \pm e_3)$ (odd number of minuses); $ \frac12 (e_5 \pm
2e_4 \pm e_1 \pm e_2 \pm e_3)$ (even number of minuses in the last
three terms) &  5 &  25 \\

\hline 36 & $(E_6,A_1\times A_1)$ & $ e_1 \pm e_2$, $ e_1 \pm
e_3$, $ e_2 \pm e_3$, $2e_1$,  $2e_2$,  $2e_3$,
$\frac{\sqrt{2}}{2} (e_4 \pm e_1 \pm e_2 \pm e_3)$ &  4 &  17 \\

\hline 37 & $(E_6,A_2)$ & $ e_1 \pm e_2$, $\sqrt{3}(e_3\pm e_1)$,
$\sqrt{3}(e_3\pm e_2)$, $2\sqrt{3}e_3$, $\frac{\sqrt{3}}{2} (e_4
\pm e_1 \pm e_2 \pm e_3)$  (odd number of minuses); $ \frac12 (e_4
\pm e_1 \pm e_2 \pm 3 e_3)$ (even number of minuses) &  4 & 15 \\

\hline 38 & $(E_6,A_1^3)$ & $ \sqrt{2}(e_1\pm e_2)$,
$2\sqrt{3}e_1$, $2e_2$, $\frac{\sqrt{2}}{2}(e_3 \pm 2 e_1 \pm
e_2)$, $e_3 \pm e_2$ &  3 & 10 \\ \hline

39 & $(E_6, A_1 \times A_2)$ & $ \sqrt{6}(e_1\pm e_2)$, $2e_1$,
$2\sqrt{3}e_2$, $\frac12(e_3 \pm 2 e_1 + 3 e_2)$,
$\frac{\sqrt{3}}2(e_3 \pm 2 e_1 - e_2)$,  $ \frac{\sqrt{2}}{2}(e_3
-3 e_2)$, $\frac{\sqrt{6}}{2}(e_3 + e_2)$ & 3 &  10 \\ \hline

40 & $(E_6,A_3)$ & $ 2(e_1\pm e_2)$, $2\sqrt{6}e_1$, $\frac12(e_3
\pm 4 e_1 + e_2)$, $e_3 \pm 2e_1 - e_2$, $ \frac{\sqrt{6}}{2}(e_3
+ e_2) $ &  3 &  8 \\ \hline


\hline

 \hline 41 &  $(H_4,A_1)$ & $e_1$, $e_2$, $e_3$,
$\frac{\sqrt{2}}2(e_1 \pm e_2 \pm e_3)$, $a e_1 \pm \frac12 e_2 \pm
b e_3$, $b e_1 \pm a e_2 \pm \frac12 e_3$, $\frac12 e_1 \pm b e_2
\pm a e_3$,  $\sqrt{2}(ae_1 \pm b e_2)$,   $\sqrt{2}(ae_2 \pm b
e_3)$,   $\sqrt{2}(b e_1 \pm a e_3)$, $\sqrt{b\sqrt{5}}(e_1\pm 2a
e_2)$, $\sqrt{b\sqrt{5}}(e_2\pm 2a e_3)$, $\sqrt{b\sqrt{5}}(e_3\pm
2a e_1)$
($a=\frac{\sqrt{5}+1}{4}$, $b=\frac{\sqrt{5}-1}{4}$) &  3 &  31 \\
\hline

\end{longtable}


There are also the following two (equivalent) families, which can be obtained by restriction of the system $F_4(\Lambda)$ related to the exceptional group $F_4.$ In section 5.3 they were denoted
$F_3^1(\Lambda)$ and $F_3^2(\Lambda)$ respectively.

\begin{longtable}{|c|p{85mm}|c|p{11mm}|}

\hline  $(F_4(\Lambda),A_1)_1$ & $e_1 \pm e_2$, $e_2 \pm e_3$,
$e_1 \pm e_3$, $\sqrt{4\Lambda^2+2}e_1$, $\sqrt{4\Lambda^2+2}e_2$,
$ \sqrt{4\Lambda^2+2}e_3$, $\Lambda\sqrt{2}(e_1\pm e_2 \pm e_3)$ &
3 &  13 \\ \hline

 $(F_4(\Lambda),A_1)_2$ & $\sqrt{2\Lambda^2+1}(e_1 \pm e_2)$,
$\sqrt{2}(e_2 \pm e_3)$, $\sqrt{2}(e_1 \pm e_3)$,
$2\sqrt{2\Lambda^2+1}e_3$, $2\Lambda e_1$, $2\Lambda e_2$,
$\Lambda(e_1\pm e_2 \pm 2 e_3)$ &  3 &  13 \\ \hline
\end{longtable}

Not all of these $\vee$-systems are different. Namely, the following
equivalences among them can be established: $$
(E_8,D_4)=F_4(\sqrt{2}), \,(E_8,D_5)= (F_4({\sqrt{2}}), A_1)_1,\,
(E_8,A_1 \times D_4)=(F_4(\sqrt{2}),A_1)_2, $$ $$ (E_7,A_1^3)_1 =
F_4(\frac12),\, (E_7,A_1^4)=(F_4(\frac12),A_1)_1, \, (E_7,A_1
\times A_3)_1=(F_4(\frac12),A_1)_2,$$
$$ (E_7, D_4)=B_3(\frac{\sqrt{2}}{2}),\, (E_6,
A_3)=B_3(-\frac23;1,1,\frac23), \, (F_4(\Lambda), A_1)_1 =
(F_4(\frac1{2\Lambda}), A_1)_2,$$
$$ (F_4(0), A_1)_1 = B_3 (0; 1,1,1) = B_3 (\sqrt 2), \, \,\, (F_4(0), A_1)_2 = B_3(-1;1,1,2), $$ where the systems
$B_n(\Lambda)$, $F_4(\Lambda)$, $B_3(\gamma; c_1, c_2, c_3)$ are
defined by \mref{lambdacl}, \mref{F4}, \mref{anc2} respectively. We can add here also the equivalence $F_4(\Lambda) =
F_4(\frac1{2\Lambda})$ (see section 5.3).

\end{document}